\documentclass[preprint,3p,times,twocolumn]{elsarticle}

\journal{Journal of \LaTeX\ Templates}
\biboptions{sort&compress}
\begin{document}

\begin{frontmatter}

\title{A generalized seniority approach to the prediction of spectroscopic factors in odd-mass Sn isotopes}
\author{Bhoomika Maheshwari\corref{mycorrespondingauthor}}
\cortext[mycorrespondingauthor]{Corresponding author}
\ead{bhoomi@um.edu.my}
\author{Hasan Abu Kassim}
\author{Norhasliza Yusof}
\address{Department of Physics, Faculty of Science, University of Malaya, 50603 Kuala Lumpur, Malaysia}
\address{Center of Theoretical and Computational Physics, Department of Physics, Faculty of Science, University of Malaya, 50603 Kuala Lumpur, Malaysia}
\author{Ashok Kumar Jain}
\address{Amity Institute of Nuclear Science and Technology, Amity University UP, Noida 201313, India} 
\author{R. F. Casten}
\address{Wright Nuclear Structure Laboratory, Yale University, New Haven, Connecticut 06520, USA}
\address{Facility for Rare Isotope Beams, Michigan State University, East Lansing, Michigan 48824, USA}
\date{\today}
\begin{abstract}

We present a study of the spectroscopic factors for the (d,p) stripping reactions to the ${11/2}^-$ states in the chain of odd-mass Sn $(A={115-131})$ isotopes by using a multi-j generalized seniority approach. The results are in line with realistic shell model calculations and explain the experimental trend quite well. The multi-j configuration used in these calculations is consistent with  earlier calculations for moments etc., and therefore, lends credence to the generalized seniority interpretation. To the best of our knowledge, this work presents the first calculation of spectroscopic factors from such an approach. 

\end{abstract}
\begin{keyword}
\texttt{Spectroscopic factors, Sn-isotopes, Generalized seniority, ${11/2}^-$ states}
\end{keyword}

\end{frontmatter}


\section{\label{sec:level1}Introduction}

The nature and evolution of the single-particle states in nuclei are of crucial importance in nuclear structure physics, and also in astrophysical processes, particularly in neutron-rich regions near magic numbers ~\cite{jones2010}, where r-processes are responsible for the production of heavier elements. A critical investigation into such astrophysical inputs can be made by means of single-particle transfer reactions, which represent a powerful tool to probe the nature of the low lying states and the energies of the single-particle orbits. In the present era of radioactive beam facilities, the study of transfer reactions in inverse kinematics has become a standard tool ~\cite{jones2010, manning2019}. The spectroscopic factors, obtained from the transfer reaction cross sections, play a key role in tracking changes in the single-particle orbits as we venture away from the stability region ~\cite{schiffer2004}.  

Various studies of spectroscopic factors in Sn isotopes have been attempted via (d,p) stripping reactions by many authors~\cite{schneid1966, cohen1968, carson1972, bingham1973, borello1975, bechara1975, fleming1982}. Our recent studies based on the generalized seniority approach ~\cite{maheshwari2016, maheshwari20161, maheshwari2017, jain2017, jain20171, maheshwari2019, maheshwari20191} have encouraged us to extend this approach to the spectroscopic factors. The present paper is the first attempt, to the best of our knowledge, to calculate the variation of spectroscopic factors in a chain of odd-mass Sn isotopes, by using the generalized seniority approach. More specifically, we focus on the spectroscopic factors from (d,p) reactions to the ${11/2}^-$ states. We have also performed realistic shell model calculations for comparison. The generalized seniority results reproduce the experimental trend quite well and lie close to the shell model results.  

The paper is divided into four sections. Section 2 defines the spectroscopic factors by using the generalized seniority approach. Section 3 presents the calculations of the spectroscopic factors for ${11/2}^-$ states in odd-mass Sn isotopes by using seniority, generalized seniority and shell model calculations. The calculated results are compared with the experimental data, wherever possible. Section 4 concludes the paper.

\section{\label{sec:level2}Spectroscopic factors}

One can define the spectroscopic amplitudes for pick-up and stripping reactions by taking the expectation values of the second-quantization creation and annihilation operators, $a^+$ and $\tilde{a}$ between the states of nuclei with $A-1$ and $A$, and $A+1$ and $A$, respectively. The corresponding spectroscopic factor may be written as

\begin{eqnarray}
SF&=&\frac{1}{2J+1} {|\langle {\Psi}^A \omega J || a^+ || {\Psi}^{A-1} {\omega}^\prime J^\prime \rangle | }^2 \nonumber \\ &=& \frac{1}{2J+1} {|\langle {\Psi}^{A-1} \omega^\prime J^\prime || \tilde{a} || {\Psi}^{A} \omega J \rangle | }^2
\end{eqnarray}
where ${a}_{m} = {(-1)}^{j-m}  \tilde{a}_{-m}$ and the factor $(2J+1)$ goes with the heavier mass $A$ by convention, and the $\omega$ indices are used to distinguish the different basis states with the same $J$ value.  

\subsection{Generalized Seniority}

The seniority scheme is generally credited to Racah~\cite{racah1943} and Flowers~\cite{flowers1952}. The complete details of seniority for a single-j shell may be found in the standard books~\cite{shalit1963,casten1990,heyde1990}. The quasi-spin scheme for identical nucleons in single-j scheme satisfies the SU(2) algebra formed by the pair creation operator $S^+$ and pair annihilation operator $S^-$, used to describe seniority. Detailed expressions and selection rules for transitions may be found in Talmi~\cite{talmi1993}. The concept of generalized seniority was first introduced by Arima and Ichimura~\cite{arima1966} for the multi-j degenerate orbits. The corresponding quasi-spin algebra for multi-j can simply be followed by defining a generalized pair creation operator $S^+ = \sum_{j} S^+_j$~\cite{talmi1993}. Talmi also introduced it for non-degenerate multi-j orbits by using $S^+ = \sum_{j} \alpha_j S^+_j$, where $\alpha_j$ are mixing coefficients~\cite{talmi1971, shlomo1972}. 

Our recent understanding of the generalized seniority approach is based on multi-j degenerate orbits by defining $S^+ = \sum_j {(-1)}^{l_j} S^+_j $, as proposed by Arvieu and Moszokowski~\cite{arvieu1966}. Here $l_j$ denotes the orbital angular momentum of the given-j orbit. This approach led us to a new set of electromagnetic selection rules and a new kind of seniority isomerism in semi-magic nuclei ~\cite{maheshwari2016}. The seniority in single-j changes to the generalized seniority $v$ in multi-j with an effective-j defined as ${\tilde{j}}= j \otimes j^\prime....$ having a pair degeneracy of $\Omega= \sum_j \frac{2j+1}{2} = \frac {(2 {\tilde{j}}+1)} {2}$. The shared occupancy in multi-j space is akin to the quasi-particle picture. However, the number of nucleons $n=\sum_j n_j$ and corresponding generalized seniority $v=\sum_j v_j$ remain integers. 

We have recently examined the goodness of seniority and generalized seniority in the behavior of the excitation energy, electromagnetic transition rates like B(EL) and B(ML) trends, $Q-$moments and magnetic moments, or g-factor values~\cite{maheshwari2016, maheshwari20161, maheshwari2017, jain2017, jain20171, maheshwari2019, maheshwari20191}. In this paper, we extend our studies to the single-particle transfers in a multi-j generalized seniority approach. The reduced matrix elements of $a^+$ in a $\tilde{j}^n$ configuration can be related to the reduced matrix elements in a $\tilde{j}^v$ configuration by using the Wigner-Eckart theorem as follows:
\begin{eqnarray}
\langle \tilde{j}^n v J || a^+ || \tilde{j}^{n-1} v-1 J' \rangle = \sqrt  { \left(\frac{2\Omega+2-n-v}{2\Omega+2-2v} \right)} \nonumber \\ \langle \tilde{j}^v v J || a^+ || \tilde{j}^{v-1} v-1 J' \rangle
\end{eqnarray} 

The above relation appears very similar to the single-j situation except for the difference in defining the multi-j pair degeneracy $\Omega= \sum_j \frac{2j+1}{2} = \frac {(2 {\tilde{j}}+1)} {2}$. We can further rewrite the Eqs. (1) and (2) as,
\begin{eqnarray}
SF &=&\frac{1}{2J+1} {| \langle \tilde{j}^n v J || a^+ || \tilde{j}^{n-1} v-1 J' \rangle | }^2   \nonumber \\
&= & v \left( \frac{2\Omega+2-n-v}{2\Omega+2-2v} \right) \nonumber \\  && \quad \quad {[ \tilde{j}^{v-1} (v-1,J') j J | \} \tilde{j}^v v J ]}^2 
\end{eqnarray}

\section{Results and Discussion}

\begin{figure}[!ht]
\begin{center}
\includegraphics[trim=55 20 80 45, clip,width=1.0\columnwidth]{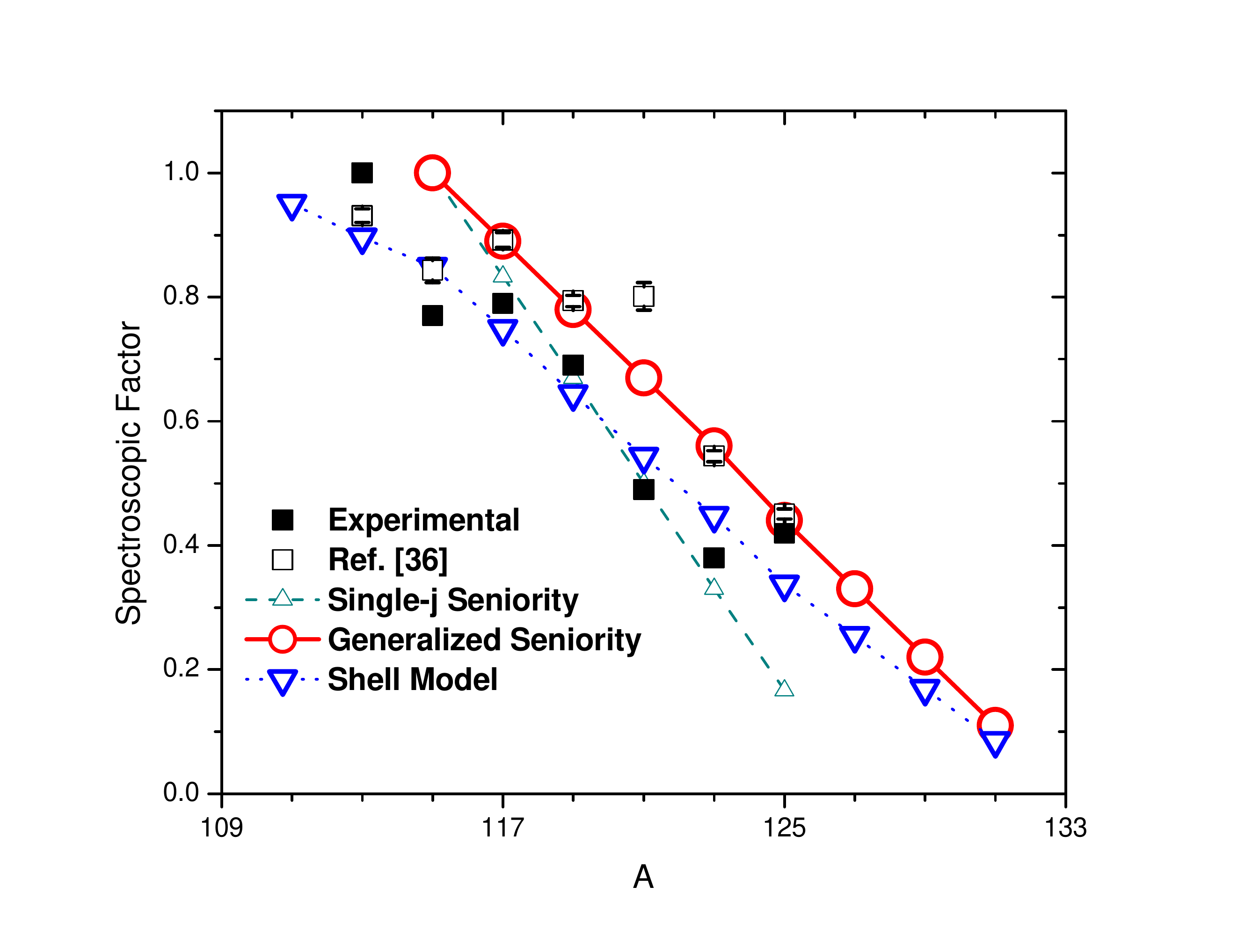}
\caption{\label{fig1}(Color online) The experimental ~\cite{blachot2009, blachot1012, blachot20091,symochko2009, ohya2010, ohya2004, katakura2010} and calculated (d,p) stripping spectroscopic factors for ${11/2}^-$ states in odd-mass Sn isotopes. The calculated values are from seniority, generalized seniority and the shell model for comparison. The generalized seniority calculations obtained by using $\Omega=9$ follow the trend, if not fully the magnitudes, of the shell model calculations. The $g_{7/2}$ orbit is fully-filled for shell model calculations due to computational limitations for $N<74$ Sn isotopes. The preliminary experimental data from the thesis~\cite{stuart} are also shown, where the quoted uncertainties are purely statistical.}
\end{center}
\end{figure}

Our main concern is to examine the multi-j configuration for the ${11/2}^-$ states in odd-mass Sn isotopes, as suggested by the generalized seniority in our previous studies of g-factor trends~\cite{maheshwari2019}. We have calculated the spectroscopic factors by using Eq.(3), for the ${11/2}^-$ states from $^{115}$Sn to $^{131}$Sn by freezing the lower-lying $g_{7/2}$ and $d_{5/2}$ orbits of the $50-82$ shell. The remaining active orbits are $h_{11/2}$, $d_{3/2}$ and $s_{1/2}$ corresponding to the pair degeneracy of $\Omega=9$. The generalized seniority has been taken as $v=1$ for these states. The coefficients of fractional parentages simply become one (from $v=0$ to $v=1$), and the trend of spectroscopic factor values can be calculated by using $\Omega =9$ and changing the neutron number $n$, as $\left( \frac{2\Omega+2-n-v}{2\Omega+2-2v} \right)$.  The generalized seniority (d,p) spectroscopic factors follow a linearly decreasing trend with increasing mass number, as shown in Fig.~\ref{fig1}, and explain the overall experimental trend. The principle experimental data for (d,p) stripping reactions in Sn isotopes have been adopted from the nuclear data sheets ~\cite{blachot2009, blachot1012, blachot20091,symochko2009, ohya2010, ohya2004, katakura2010}. It may be noted that the adopted data sets for (d,p) stripping reactions in these isotopes have not changed much since the previous evaluations during 1980-90s. Our generalized seniority calculated results are in line with the previous interpretations of pairing correlations in the Sn isotopes by Andreozzi $et$ $al.$ ~\cite{andreozzi1996}, where they have shown the dominance of $h_{11/2}$, $d_{3/2}$ and $s_{1/2}$ orbits above $N=64$. Table ~\ref{tab:table1} compares the spectroscopic factors obtained from our generalized seniority results and the previous results of Andreozzi $et$ $al.$ ~\cite{andreozzi1996}. Besides, preliminary experimental data from a Ph.D. thesis ~\cite{stuart} have been shown for comparison in Table~\ref{tab:table1} and Fig.~\ref{fig1}.   

\begin{table}[htb]
\caption{\label{tab:table1}A comparison of the experimental and calculated (d,p) stripping spectroscopic factor values for the ${11/2}^-$ states in $N>64 $ odd-mass Sn isotopes. The principle experimental data have been taken from nuclear data sheets~\cite{blachot2009, blachot1012, blachot20091,symochko2009, ohya2010, ohya2004, katakura2010} for comparison. GS denotes the generalized seniority results, which have also been compared to the previous studies by using a pairing Hamiltonian approach \cite{andreozzi1996}. The last column presents the preliminary experimental data taken from the thesis~\cite{stuart}, where the quoted uncertainties are purely statistical.}
\begin{tabular}{c c c c c}
\hline
Nucleus & Exp. & GS (ours) & Ref.~\cite{andreozzi1996} & Ref.~\cite{stuart} \\
\hline
&&&&\\
$^{115}$Sn & 0.77 & 1.00 & 0.89 & 0.843(20)\\
$^{117}$Sn & 0.79 & 0.89 & 0.83 & 0.892(12)\\
$^{119}$Sn & 0.69 & 0.78 & 0.70 & 0.794(9)\\
$^{121}$Sn & 0.49 & 0.67 & 0.61 & 0.801(22)\\
$^{123}$Sn & 0.38 & 0.56 & 0.52 & 0.544(9)\\
$^{125}$Sn & 0.42 & 0.44 & 0.42 & 0.451(8)\\
$^{127}$Sn &  & 0.33 &  &\\
$^{129}$Sn &  & 0.22 &  &\\
$^{131}$Sn &  & 0.11 &  &\\
\hline
\end{tabular}
\end{table}

\begin{table*}[!htb]
\caption{\label{tab:table2}A list of calculated occupancies of various orbits for the ground-state in $N>63$ even-mass Sn isotopes, from the shell model and the generalized seniority approach. The $g_{7/2}$ and $d_{5/2}$ orbits are assumed to be fully occupied in the generalized seniority approach.}
\begin{center}
\begin{tabular}{c c c c c c c c c c c}
\hline
Nucleus & \multicolumn{5}{c}{Shell Model} & \multicolumn{5}{c}{Generalized Seniority} \\
\hline 
& $g_{7/2}$ & $d_{5/2}$ & $d_{3/2}$ & $s_{1/2}$ & $h_{11/2}$ \quad  & \quad $g_{7/2}$ & $d_{5/2}$ & $d_{3/2}$ & $s_{1/2}$ & $h_{11/2}$ \\
\hline
&&&\\
$^{114}$Sn & 8.0 & 4.25 & 0.26 &  0.12 & 1.37 \quad & \quad 8.0 & 6.0 & 0.0 & 0.0 & 0.0  \\
$^{116}$Sn & 8.0 & 4.52 & 0.47 & 0.27  & 2.75 \quad & \quad 8.0 & 6.0 & 0.44 & 0.22 & 1.32 \\
$^{118}$Sn & 8.0 & 4.71 & 0.75 & 0.44 & 4.09 \quad & \quad 8.0 & 6.0 & 0.88 & 0.44 & 2.64 \\
$^{120}$Sn & 8.0 & 4.91 & 1.12 & 0.64 & 5.33 \quad & \quad 8.0 & 6.0 & 1.32 & 0.66 & 3.96  \\
$^{122}$Sn & 8.0 & 5.11 &  1.53 & 0.85 & 6.52 \quad & \quad 8.0 & 6.0 & 1.76 & 0.88 & 5.28  \\
$^{124}$Sn & 7.32 & 5.31 & 2.22 & 1.26 & 7.89 \quad & \quad 8.0 & 6.0 & 2.24 & 1.12 & 6.72  \\
$^{126}$Sn & 7.56 & 5.49 &  2.57 & 1.46 & 8.92 \quad & \quad 8.0 & 6.0 & 2.68 & 1.34 & 8.04  \\
$^{128}$Sn & 7.74 & 5.66 &  2.99 & 1.65 & 9.97 \quad & \quad 8.0 & 6.0 & 3.12 & 1.56 & 9.36  \\
$^{130}$Sn & 7.88 & 5.83 & 3.50 & 1.82 & 10.98 \quad & \quad 8.0 & 6.0 & 3.56 & 1.78 & 10.68 \\
\hline
\end{tabular}
\end{center}
\end{table*}

To further support our results, we have performed the shell model calculations by using the Sn100PN interaction ~\cite{brown05}. The interaction assumes $^{100}$Sn as a core. The neutron single-particle energies are taken to be $-10.6089, -10.2893, -8.7167, -8.6944, -8.8152$ MeV for $1g_{7/2}$, $2d_{5/2}$, $2d_{3/2}$, $3s_{1/2}$ and $1h_{11/2}$ orbits, respectively. The two-body matrix elements are employed with the mass dependence of $A^{-1/3}$. The Nushell code~\cite{brown} was used to diagonalize the shell model Hamiltonian and evaluate the spectroscopic factors. The calculations were truncated for $N<74$ Sn isotopes by freezing the $g_{7/2}$ orbit due to computational difficulty in handling large dimensions.  However, the results should largely remain unaffected, as only $h_{11/2}$, $d_{3/2}$ and $s_{1/2}$ orbits are needed for studying the ${11/2}^-$ states, as suggested by the generalized seniority approach.     

The shell model calculated results are also shown in Fig.~\ref{fig1}. Interestingly, the overall trend from the shell model follows the generalized seniority calculated results. Such investigations, hence, validate the usage of generalized seniority and corresponding multi-j configuration for explaining the ${11/2}^-$ states in odd-mass Sn isotopes. We have also plotted the single-j seniority trend obtained from a single-j $h_{11/2}$ configuration in Fig.~\ref{fig1}. As may be noticed, the single-j approach cannot explain the full trend. The seniority trend exhibits a different slope from the generalized seniority and the shell model trends. This may be due to the consideration of various orbits in generalized seniority so that the $h_{11/2}$ occupancy increases slower with increasing neutron number, in comparison to the pure seniority case. The goodness of generalized seniority is hence crucial for explaining the spectroscopic properties in Sn isotopes. The success of this simple model suggests that the large scale shell model calculations can also be simplified by using a limited space without sacrificing the results. The generalized seniority approach also gives us a simple physics understanding of the spectroscopic factors. 

We have also calculated the occupancies of the $h_{11/2}$, $d_{3/2}$ and $s_{1/2}$ orbits for the ground states of even-mass Sn isotopes from the generalized seniority approach by using Eq.(3) for (d,p) stripping reactions in Sn isotopes, where the $g_{7/2}$ and $d_{5/2}$ orbits are taken to be fully-filled. The calculated numbers are also compared with the shell model calculations, as listed in Table~\ref{tab:table2}. The numbers from both the models do not differ much from each other for the $h_{11/2}$, $d_{3/2}$ and $s_{1/2}$ orbits, particularly for neutron-rich Sn isotopes. 
Hence, the multi-j $h_{11/2} \otimes d_{3/2} \otimes s_{1/2}$ configuration corresponding to $\Omega=9$ in a generalized seniority approach works pretty well for neutron-rich Sn isotopes and also supports the sub-shell gap at $N=64$.

\section{\label{sec:level3}Conclusion}

In this paper, we have carried out a study of the spectroscopic factors for the (d,p) stripping reactions populating the ${11/2}^-$ states in the Sn isotopes by using the multi-j generalized seniority approach. The calculated spectroscopic factors follow the experimental trend quite well. Although, the importance of pairing effects in these nuclei has long been recognized, we believe that our work supports the goodness of generalized seniority and also validates our previous interpretation of various other spectroscopic properties like B(EL) trends, half-lives, high-spin isomers, g-factor trends etc. in terms of generalized seniority. It is pertinent to note the remarkable consistency of the same multi-j configurations in explaining all the spectroscopic properties of Sn isotopes.

The generalized seniority results also follow the overall trend of realistic shell model calculations, which are a bit challenging computationally due to the large dimensions. The generalized seniority approach thus turns out to be a simple way to obtain the spectroscopic factors, which also provides an understanding of the underlying physics. The explanation, in a way, also supports the continued existence of the sub-shell gap at $N = 64$ for Sn isotopes. The role of symmetries in terms of the goodness of generalized seniority is, therefore, crucial in explaining the spectroscopic properties of the Sn isotopes. We propose to expand these studies to other mass regions also. 
 
\section*{Acknowledgements}
We thank Sean J. Freeman, University of Manchester, for clarifying the preliminary nature of data from ~\cite{stuart}. BM gratefully acknowledges the financial support in the form of a Post-Doctoral Research Fellowship from University of Malaya, Kuala Lumpur. HAK and NY acknowledge support from the Research University Grant (GP0448-2018) under University of Malaya. AKJ would like to thank Amity University for research facilities to continue this work.


\end{document}